\documentclass[3p]{elsarticle}
\usepackage[numbers]{natbib}
\usepackage{subcaption}
\usepackage{graphicx}
\usepackage{float}
\usepackage{amsmath}
\usepackage{lineno}
\usepackage[frozencache=true,cachedir=.]{minted}

\usepackage{textgreek}
\usepackage{xspace}
\usepackage{setspace}

\newcommand{\tripoli}{{\texttt{Tripoli-4}}\textsuperscript{\textregistered}\xspace}
\newcommand{\mcnp}{\texttt{MCNP}\xspace}
\newcommand{\fifrelin}{\texttt{FIFRELIN}\xspace}
\newcommand{\geant}{\texttt{GEANT4}\xspace}
\newcommand{\toucans}{\texttt{TOUCANS}\xspace}
\newcommand{\ncrystal}{\texttt{NCrystal}\xspace}

\newcommand{\funz}{\texttt{FUNZ}\xspace}
\newcommand{\cadmesh}{\texttt{CADMesh}\xspace}
\newcommand{\fispact}{\texttt{FISPACT}\xspace}

\journal{Nucl. Inst. and Meth. in Phys. Res. Sec. A}

\begin{document}

\begin{frontmatter}

\let\WriteBookmarks\relax
\def\floatpagepagefraction{1}
\def\textpagefraction{.001}

\title{\toucans : a versatile Monte Carlo neutron transport code based on \geant.} 

\author[1]{L. Thulliez}\corref{cor1}
\ead{loic.thulliez@cea.fr}
\author[1]{B. Mom}
\author[1]{E. Dumonteil}

\address[1]{Université Paris-Saclay, CEA, Institut de Recherche sur les Lois Fondamentales de l'Univers, 91191, Gif-sur-Yvette, France}

\cortext[cor1]{Corresponding author}

\begin{abstract}
This paper reports the development of \toucans, a new Monte Carlo neutron transport code fully written using the \geant toolkit. It aims at modeling complex systems easily and rapidly, thanks to a simple key-value input file. While its main focus is the transport of low energy neutrons -below 20 MeV- using evaluated cross-section libraries, it also takes advantage of the state-of-the-art capabilities offered by \geant regarding the modeled cross-sections for practically all particles at all energies. Hence it gives access to non-specialists to a wide variety of applications involving different particles, such as for instance the simulation of compact accelerator based neutron sources (CANS) which initially motivated its development. It leverages on recent implementations performed within \geant to provide both shielding calculation capabilities (using Adaptive Multilevel Splitting, which is a fully automated variance reduction technique) and criticality calculation capabilities (to estimate the $k_{\text{eff}}$ of a multiplicative system). It also benefits from a tight coupling with \cadmesh to readily import CAD models, with \ncrystal to extend the support of thermal neutron transport to crystals, with \fifrelin to increase compound nucleus de-excitation accuracy, and with \funz to provide to the user parametric simulation capabilities (such as multi-objective algorithms, calculation of Pareto sets, etc). Thanks to recent improvements of the neutron physics within the \geant Neutron-HP package, the \toucans code can be considered on-par with industrial neutron transport codes such as \tripoli or \mcnp in term of neutron physics accuracy. Its sources should be released soon amongst \geant advanced examples.
\end{abstract}



\end{frontmatter}


\section{Introduction}

The open-source Monte Carlo toolkit \geant developed by a CERN collaboration~\cite{Allison2016} has a wide community of users ranging from fundamental high energy physics to medical and space applications. 
On the side of low-energy neutron transport (below 20 MeV), its use in the past was limited due to the specific handling of evaluated cross-sections libraries, conjugated to prohibitive calculation times (the tracking steps being common to all particles) and to the initial time required to develop a specific application using the \geant C++ toolkit. Also, industrial-grade Monte Carlo neutron transport codes targeting the modeling of nuclear reactors were already providing such capabilities -as well as many others- hence discouraging the development of up-to-date evaluated neutron physics within the toolkit.
However, the past decade has seen an increase in the use of thermal and cold neutron sources for fundamental physics (such as in~\cite{nesvizhevsky_quantum_2002}) 
and for condensed or soft matter experiments (see for instance~\cite{IAEA_CANS_2021}). Furthermore the shutdown of many experimental reactors, providing neutrons as a probe for condensed matter studies, increases the need to develop dedicated facilities and to study new concepts such as neutron source amplifiers using a fissile medium~\cite{galy_neutron_2002} or compact accelerator based neutron sources, also known as CANS. The development of such concepts however requires dedicated algorithms and physics. 
In the case of neutron source amplifiers for instance, the so-called power iteration method allows to perform a criticality-safety study of the system through the computation of its multiplication factor (the so-called $k_{\text{eff}}$). 
Regarding the design of a CANS, such as the SONATE facility currently developed at CEA-Saclay (France)~\cite{ott_frederic_sonate_2020}, the transport code needs to be able to simulate an even wider range of physics apart from the neutron transportation. In fact the neutrons are produced through $^{7}$Li(p,n) or $^{9}$Be(p,n) reactions and are transported, using variance reduction techniques, to a detector in which the electromagnetic processes have to be modelled.
All these requirements motivated both the improvements of \geant Neutron-HP package models, as well as the development of \toucans, a proper Monte Carlo neutron transport code providing all these functionalities and algorithms to its users through a versatile easy-to-use key-value input file, avoiding the need to recompile the whole \geant application at each parametric study. \toucans originally stands for 'TOolkit for Unified CANS design' but has now extended its range of applications to the study of diverse neutron transport applications. While its neutron physics accuracy is now close to industrial transport codes such as \tripoli~\cite{tripoli4} or \mcnp~\cite{mcnp6}, this code also allows to transport accurately all other particles (gammas, electrons, positrons, ions, etc.) necessary to simulate various physics and situations, as for instance nuclear detection systems, nuclear instrumentation, criticality-safety studies or shielding studies. This code has already been used successfully in numerous studies such as in refs~\cite{Sonate_Thulliez2020,Thulliez2021,Hirtz2022,Mom2022}.
This paper presents the main features of \toucans, while giving examples of its use through the specific case of the modeling of a neutron source whenever possible. It is structured as follows. In Section \ref{sec:toucans_features}, the main functionalities of the code are reviewed, while Section \ref{sec:toucans_framework} presents its user interface. The low energy neutron physics of the code is presented in Section \ref{sec:toucans_physics}. Then Sections \ref{sec:toucans_generator} and \ref{sec:toucans_algo} respectively detail the primary event generator and the diverse \geant algorithms that can be used within \toucans. Finally Section \ref{sec:toucans_coupling} gives an example of its coupling to a multi-objective optimization software.

\section{Main functionalities of \toucans}
\label{sec:toucans_features}

The key features of the \toucans code are summarized below and will be detailed in the next sections:
\begin{itemize}
    \item \textbf{Input file format with key-value} structure allowing a straightforward definition of the setup geometry elements and their properties (position, dimension, material, scorers, etc). 

    \item \textbf{CAD geometry} import with the \cadmesh software~\cite{poole2012acad, poole2012fast}.
    
    \item \textbf{Accurate neutron interactions}: thermal scattering laws with up-to-date evaluated data (ENDF/BVIII.0 and JEFF-3.3)~\cite{THULLIEZ2022} and \ncrystal~\cite{caiKittelmann2020,KittelmannCai2021}, Doppler broadening rejection correction (DBRC)~\cite{Marek2023a}, probability tables for the treatment of the unresolved resonance region~\cite{Marek2023b}, accurate description of compound nuclear reaction with \fifrelin~\cite{Litaize2015}.    

    \item \textbf{Built-in neutron tallies/scorers} such as track-length and collision flux estimators, energy deposition, 2D or 3D flux maps for shielding applications, fission matrix, etc. Possibility to easily implement any user-defined tallies. 
    
    \item \textbf{Versatile primary event generator} that can be defined through multiple formats (\textit{e.g.} text file, ROOT TH1 or TH2~\cite{ROOT}, etc) allowing to tailor the characteristics of the primary events as close as possible to the reality (energy-angle correlation, spatial distribution coming from a charged particle beam interacting with a target, etc). A set of neutron source terms are available such as $^{7}$Li(p,n), $^{9}$Be(p,n) or fission source such as $^{252}$Cf(sf) and $^{235}$U(n$_{\text{th}}$,f).
       
     \item \textbf{Multiplicative factor ($k_{\text{eff}}$) estimation} for criticality-safety studies, using the power iteration method with the combing algorithm~\cite{Booth1996ACarlo}.

    \item \textbf{Adaptive Multilevel Splitting} algorithm~\cite{Cerou2007} which is a plug-and-play variance reduction technique to study high attenuation problems encountered in shielding applications such as in radio-protection~\cite{Louvin2017} or low background experiments (dark matter or neutrino studies).
     
    \item \textbf{Easy coupling to external codes} such as multi-objective optimization algorithms to find the set of optimal solutions for an experimental setup design with requirements (\textit{e.g.} with \funz software~\cite{Richet_Funz2021}), or nuclear element evolution codes solving the Bateman equation (\textit{e.g.} \fispact~\cite{Sublet2017}).
\end{itemize}

\section{\toucans input file}
\label{sec:toucans_framework}

Since \toucans is fully relying on the \geant toolkit, it inherits from its many functionalities, ranging from its geometry visualization tools to its wide physics lists, and also takes advantage of the latest add-on tools developed by community members. However, such a broad number of capabilities comes at the price of a strong initial investment in the building of a specific \geant application. Therefore, \toucans's philosophy consists in releasing to a broad community of non-specialist users the vast physics options offered by \geant, with a particular emphasis on low energy neutron physics, through a simple user interface that allows to easily describe the setup to simulate. An example of a full input file is presented in Appendix \ref{annexe:inputCard} and will be detailed throughout the paper, together with many of other available functionalities. 

The syntax of the \toucans input file is presented Figure~\ref{fig:toucansMainBlock}, for the particular example of the geometry declaration. The key-value data are read if enclosed within a dollar sign \$, otherwise they are discarded. It has to be pointed out that this input file key-value structure is similar to the one used in \fifrelin~\cite{Litaize2015, FifrelinUserGuide}. Internally the code interprets the instructions and performs all the steps needed by \geant through generic methods, in this example, to get a \geant physical volume. The advantage to have such an input file lies in the fact that all the parameters are now accessible externally while masking the complex underlying internal structure of \geant C++ objects and no code needs to be recompiled. This approach is therefore similar in spirit to the use of a script using the \geant user interface. It lends itself more easily to a coupling with external dedicated algorithms such as search-grid, multi-objective or genetic algorithms, multi-physics phenomena, etc.

\begin{figure}[ht] 
\begin{minted}[fontsize=\small]{C++}
$ STRING      MyVolume/MotherVolume   my_motherVolume $ 
$ STRING      MyVolume/Type           my_type         $
  [BOX, CYLINDER, SPHERE, POLYCONE, etc, STL-CADMesh]
$ DOUBLELIST  MyVolume/Dimensions     dimX dimY dimZ  $
  [set the one related to the defined volume in cm]
$ DOUBLELIST  MyVolume/Position       posX posY posZ  $
  [position in the mother volume frame in cm]
$ STRINGLIST  MyVolume/Rotation/Axis  my_axis         $
  [X, Y, Z]
$ DOUBLELIST  MyVolume/Rotation/Angle my_angle        $
  [in degree]
$ STRING      MyVolume/Material       my_material     $ 
  [material from GEANT4 or NCRYSTAL]
$ STRING      MyVolume/Color          my_color        $ 
  [blue, red, green, yellow, etc]
$ STRINGLIST  MyVolume/Scorers        my_scorer       $ 
  [user defined scorers handled by a scorer manager
   CELL_FLUX, TTREE_RECORDER, etc]
$ STRINGLIST  MyVolume/Subtract/List  my_minusVolume  $
  [list of volume to subtract to MyVolume]
$ STRINGLIST  MyVolume/Union/List     my_plusVolume   $
  [list of volume to add to MyVolume]
\end{minted}
\caption{\toucans building block inside an input file.}
\label{fig:toucansMainBlock}
\end{figure} 

\begin{figure}[htbp]
    \begin{center}
	\includegraphics[scale=0.15]{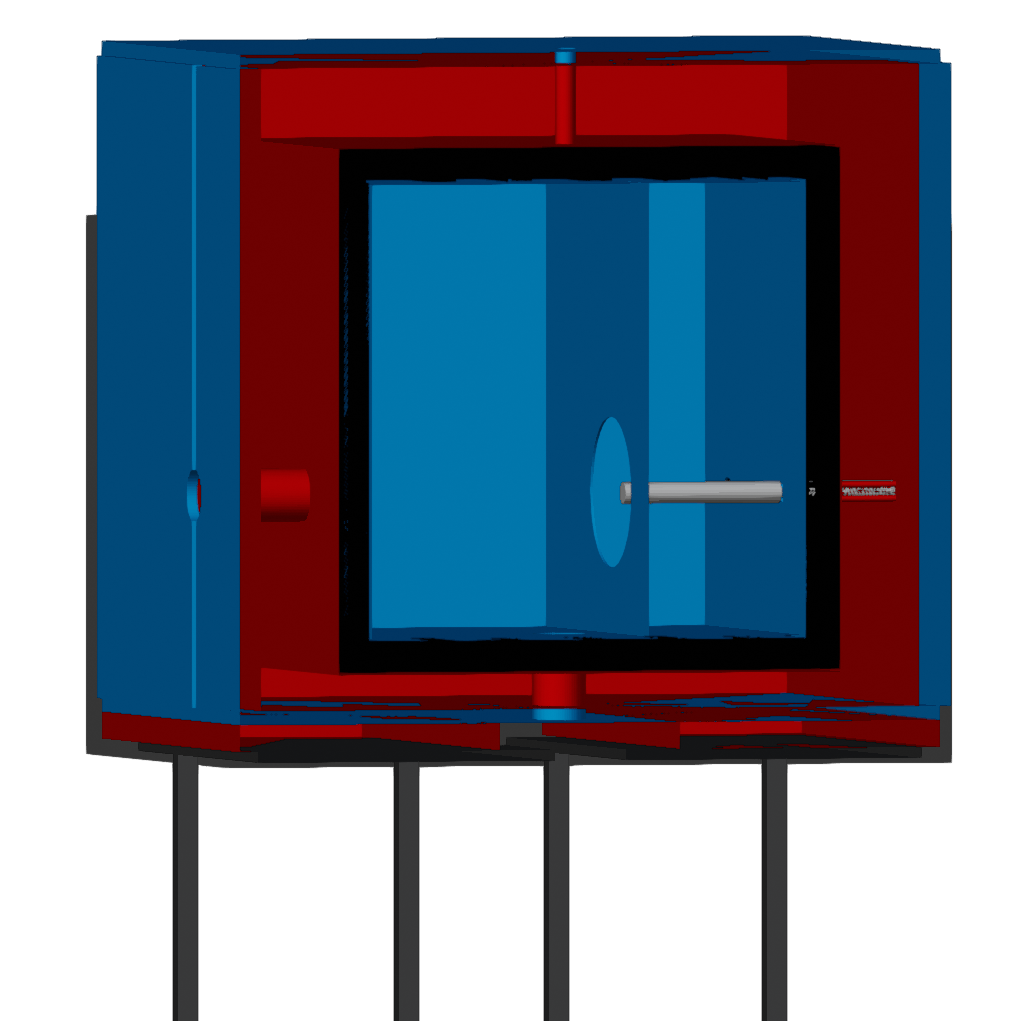} 
	\end{center}
	\caption{\label{fig:CADvisu} CAD visualization of target-moderator-shielding assembly of a CANS, sectional view. The geometry is imported \textit{via} \cadmesh in \geant. The rendering is done with Blender software \cite{BlenderSoftware} \textit{via} VRML format.}
\end{figure}

\noindent
A particular care has been devoted to the support of CAD geometry imports, as this feature is key to translating industrial projects (described by CAD software in STL format for example) to Monte Carlo simulations. In \geant this can be done through the \cadmesh add-on~\cite{poole2012acad, poole2012fast}. Such an example of a CAD support within \toucans is given Figure~\ref{fig:CADvisu}, where the import of a CANS target-moderator-shielding assembly is represented. This feature helps to identify any difficulties which could rise when mechanical constraints are set to actually build a facility/setup and lead to performance degradation or neutron activation problems, etc. If the CAD volumes are all generated in the same frame and well positioned relative to the others, \geant will place accurately the volumes relative to each other in the space. In \toucans, the CAD import is done with the following instructions:
\begin{minted}[fontsize=\small]{C++}
$ STRING      MyVolume/Type           STL         $
$ STRING      MyVolume/STL/FilePath   my_stl_path $
\end{minted}
\noindent It can be difficult to avoid overlap problem (that could lead to particle tracking problems) in \geant because of numerical precision, meaning that often there is volume overlapping by just few femtometers. To avoid this, the volume being defined has to be subtracted to its mother volume. 

\section{Low energy nuclear physics and neutron transport}
\label{sec:toucans_physics}

\subsection{Verification and validation of the low energy neutron transport physics}

\begin{figure}[htbp]
    \begin{center}
	\includegraphics[scale=0.4]{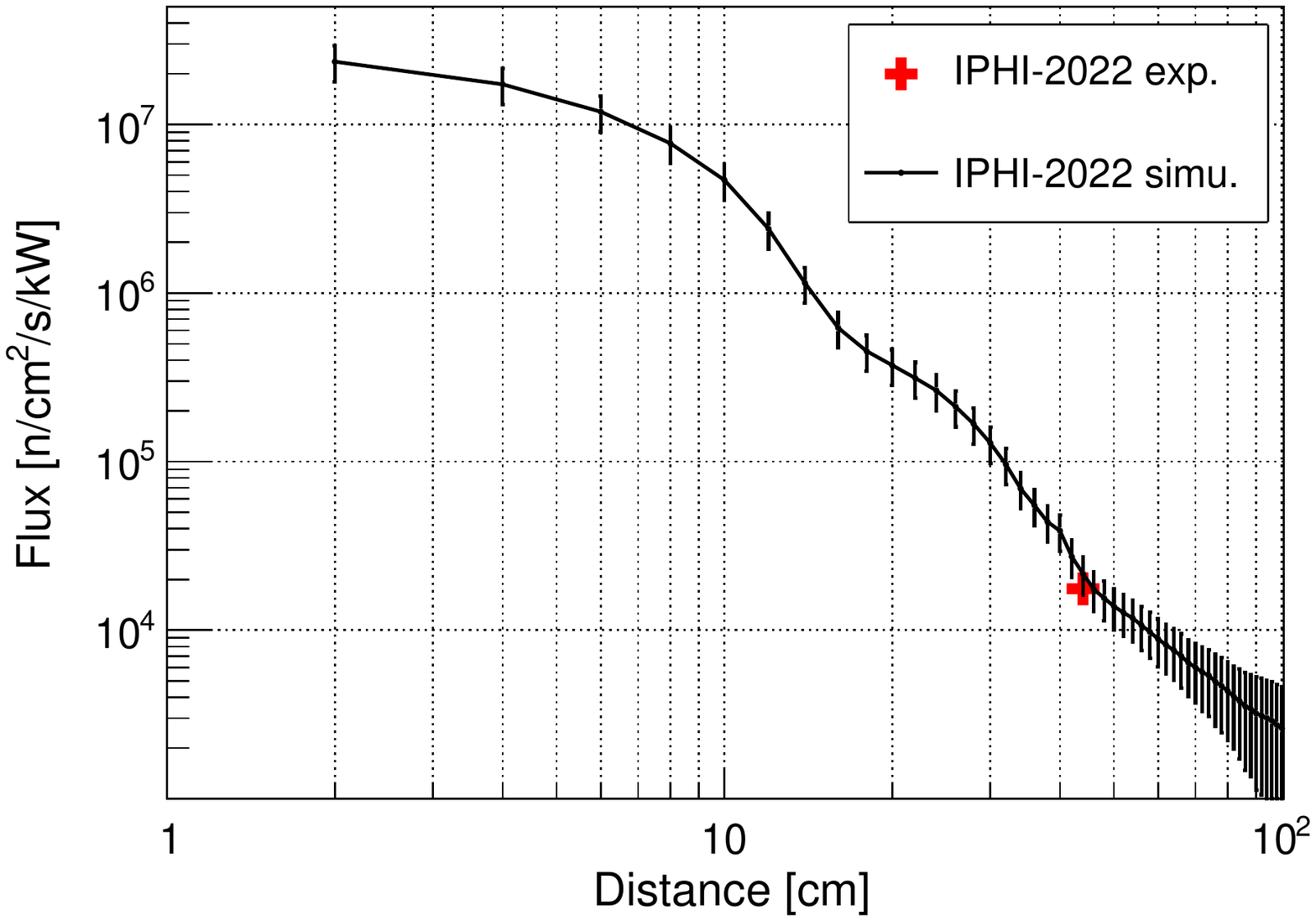} 
	\end{center}
	\caption{\label{fig:TOUCANSvalidationAgainst2022Exp} Comparison between \toucans thermal neutron flux (E~$\le$~512 meV) prediction and Au and Au+Cd dosimeter measurements allowing to determine the neutron flux below 512 meV (cadmium cut) performed in the 2022 CANS configuration at the IPHI-accelerator (CEA-Saclay).}
\end{figure}

The initial motivation of \toucans was to provide a simple code to model specifically low energy neutron transport applications (below 20 MeV), as its first use was devoted to the design and study of a CANS facility. This objective could be met thanks to important improvements of the \geant Neutron-HP package the past few years~\cite{Mendoza2014,Mendoza2018,Tran2018a, THULLIEZ2022, Marek2023a, Marek2023b}, making the neutron physics accuracy now on-par with reference neutron transport codes such as \tripoli and \mcnp. These latest developments are: the treatment of thermal neutron with the SVT algorithm and the up-to-date evaluated thermal scattering laws (TSL) (ENDF/B-VII.1, ENDF/B-VIII.0, JEFF-3.3)~\cite{THULLIEZ2022}, the implementation of the Doppler Broadening Rejection Correction (DBRC) algorithm~\cite{Becker2009,Marek2023a} and the possibility to use probability tables to describe the unresolved resonance region~\cite{Marek2023b}.
\newline
\indent
All the aforementioned developments were verified and validated against reference neutron transport codes such as \tripoli or \mcnp with dedicated scripts which are now part of the \geant toolkit. \toucans and so the Neutron-HP package have also been validated against different CANS experimental configurations tested at the 3 MeV proton IPHI accelerator~\cite{Senee2018} with beryllium targets since 2016 at CEA-Saclay (France). These different validations were performed on different CANS configurations in 2016~\cite{Sonate_Tran_2020}, in 2019~\cite{Sonate_Thulliez2020} and in 2022~\cite{Mom2022,Schwindling2022}. They have shown that numerical data produced by \geant were in agreement with these CANS experimental data. The latest validation, \textit{i.e.} for the 2022 configuration, is shown in Figure~\ref{fig:TOUCANSvalidationAgainst2022Exp} where the neutron flux versus distance to the production vertex experimentally measured at IPHI is in good agreement with the \toucans simulated data.

\subsection{Thermal neutron scattering with \ncrystal}
To complete the thermal scattering data from the standard evaluated libraries, the \ncrystal library can be used in \geant through its simple interface/hook. It contains more than 130 materials in its latest version (ncrystal-3.5.1)~\cite{caiKittelmann2020,KittelmannCai2021}. For example, in CANS development it can be used to simulate the response of single-crystals playing the role of low pass thermal neutron filters. In \toucans, \ncrystal materials can be added easily \textit{via} the following key-values with the possibility to tune the crystal properties:
\begin{minted}[fontsize=\small]{C++}
$ STRING      MyVolume/Material/Type      NCRYSTAL    $ 
  [NCRSYTAL, GEANT4 (default)]
$ STRINGLIST  MyVolume/Material/Options   nc_options  $
  [PHONON_DEBYE, BRAGG, NO_BRAGG]
$ STRING      MyVolume/Material           nc_material $ 
\end{minted}

\subsection{Compound nucleus reaction with \fifrelin}
To improve the accuracy of \geant regarding the neutron, photon and conversion electron emissions of a compound nucleus produced either by neutron capture or fission (\textit{i.e.} the primary fission fragments), the possibility to use the \fifrelin code~\cite{Litaize2015} has been added to \toucans. \fifrelin is a de-excitation Monte-Carlo code using the Hauser-Feshbach framework with up-to-date neutron transmission coefficient, radiative strength function and level density models as described in~\cite{RIPL3}. It also uses RIPL-3 \cite{RIPL3} and EGAF \cite{EGAFpubli} evaluated databases to take into account all the experimental nuclear level information. The impact of these nuclear models on the results can so be studied. The energy, spin and parity conservation and transition selection rules to go from one initial to a final nuclear state are intrinsically taken into account. Therefore the use of \fifrelin is really an advantage compared to the use of an evaluated library in which only the energy and particle multiplicity are conserved on average. This could be of interest for homeland security to insure a non-proliferation policy based on advanced nuclear material detectors or for precise physics experiments. In fact the coupling between \fifrelin and \geant has already shown its ability to precisely describe these reactions in the context of fundamental (neutrino) physics~\cite{Almazan2019,Thulliez2021,crab2022}. 
To use \fifrelin within \toucans, the following \geant environment variable has to be set: \textit{GEANT4\_FIFRELIN\_DATA="path\_to\_fifrelin\_database"}. The coupling is then performed through the following commands:

\begin{minted}[fontsize=\small]{C++} 
$ BOOL        PhysicsList/Fifrelin/IsON       1  $
    [0=disable (default) or 1=enable]
$ INTLIST     PhysicsList/Fifrelin/Isotopes   ZA $
    [with ZA=Z*1000+A / multiple isotopes can be set ]
    [isotope data have to be in GEANT4_FIFRELIN_DATA]
\end{minted}

\section{Primary event generators} 
\label{sec:toucans_generator}
\begin{figure}[htbp]
    \begin{center}
	\includegraphics[scale=0.4]{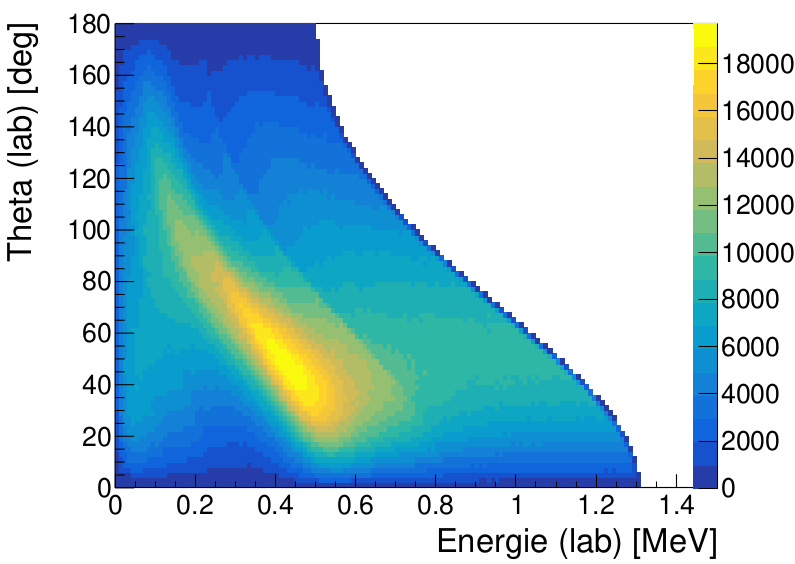}
	\end{center}
	\caption{\label{fig:neutronSource_7Li_p3MeV} Properties of the neutrons produced by the reaction p(3MeV)+$^{7}$Li $\rightarrow$ n+$^{7}$Be, in the laboratory frame based on~\cite{Liskien1975,Herrera2014} as explained in the text. This TH2F can be easily read by \toucans with key="Generator/Type" set to value="ENERGY\_ANGLE\_CORRELATION".}
\end{figure}

The declaration of the primary neutron source requires to define the energy, the momentum and the position of the emitted neutrons in the input file. The primary event generator changes slightly in philosophy if one deals with a fixed neutron source (\textit{e.g.} CANS) or with a generational neutrons source (\textit{e.g.} neutron transport in multiplicative media). For this latter, indeed, the initial position of the neutrons does not matter much since the neutrons have to spatially converge to the eigenvalue of the problem in using the power iteration method (cf. Section \ref{subsec:toucans_algo_criticality}). Therefore a convergence criterion has to be reached before scoring (this criteria can be applied either to the $k_{\text{eff}}$ or to the entropy of the neutron source~\cite{Lux1991,Ueki,Nowak2016}). For neutrons produced by fission reactions, either a Watt spectrum can be used or if more precise data are needed for example taking into account neutron/gamma correlation, dedicated codes such as \fifrelin can be used.
\newline \indent
For fixed neutron source applications, such as CANS, the initial position is directly given by the charged particle or neutron beam profile on the target converter or by the localisation of the primary neutron source. To determine the properties of the neutrons produced by a charged particle beam on a target converter (yield, energy, momentum, etc), different codes can be used depending on the beam energy. For example, for nuclear reactions induced by protons or deutons of few MeV, the DROSG code can provide differential cross-sections for the main target used in the laboratory (d, t, Li, Be, B, etc)~\cite{Drosg2000}. For 10-100 MeV beam energies, often theoretical approaches fail because it is hard to take into account in the same framework both the structure of a nucleus and the dynamics of the reactions. In this case the TENDL evaluated library~\cite{Koning2019} can be used. At higher energies, such as for spallation reactions, intra-nuclear cascade codes such as INCL~\cite{Mancusi2014, Hirtz2020} are recommended. 
To illustrate further the use of the \toucans code in the development of a CANS, we now give the example of the $^{7}$Li(p,n) reaction used in boron neutron capture therapy~\cite{Herrera2014}. Between 1.95 MeV and 7 MeV the Liskien and Paulsen evaluation data can be used~\cite{Liskien1975}. This evaluation was extended to the threshold energy, \textit{i.e.} from 1.95 MeV to 1.88 MeV, using the Breit-Wigner one-level formalism as detailed in~\cite{Herrera2014}. The results are presented in Figure~\ref{fig:neutronSource_7Li_p3MeV} and stored in a TH2F structure (ROOT-CERN) \cite{ROOT}, which can be directly provided to \toucans \textit{via} the following command lines:
\newline

\begin{minted}[fontsize=\small]{C++}
$ STRING        Generator/Type                  my_type     $
    [FULL_USER, FIFRELIN, TTREE, ENERGY_ANGLE_CORRELATION]
$ STRING        Generator/PositionSampler       my_sampler  $ 
    [GAUSSIAN, UNIFORM_DISK, UNIFORM_BOX, TH2F, TTREE]
    
# in the following trunk=Generator/PositionSampler
$ DOUBLELIST    trunk/Global/Position           x  y z      $ 
$ STRINGLIST    trunk/Global/Rotation/Axis      my_axis     $ 
$ DOUBLELIST    trunk/Global/Rotation/Angle     my_angle    $ 
  
//IF my_type==ENERGY_ANGLE_CORRELATION:
  $ INT         Generator/PDGCode               my_pdgCode  $ 
  $ STRING      Generator/EnergyAngleSampler    TH2F        $ 
      [TEXT, TH2F]
  # in the following trunk=Generator/EnergyAngleSampler
  $ STRING      trunk/TH2F/Path             root_file_path  $ 
  $ STRING      trunk/TH2F/Name                 th2h_name   $ 
  //OR
  $ STRING      trunk/Text/dXSOverdTheta    root_file_path  $ 
  $ STRING      trunk/Text/ddXSOverdThetadE     th2_name    $ 
  
//ELSE IF my_type==FULL_USER:
  $ STRING        Generator/EnergySampler       my_sampler  $
    [FIXED, WATT_SPECTRUM, TH1F, TEXT, TENDL]
  $ STRING        Generator/MomentumSampler     my_sampler  $ 
    [RESTRICTED_THETA, ISOTROPE]

//ELSE IF my_type==FIFRELIN:
  $ STRING  Generator/Fifrelin/Path             my_path     $
    [read the FIFRELIN file including pdgCode, etc]
    [if not provided define Generator/MomentumSampler]
    
//ELSE IF my_type==TTREE:
  $ STRING  Generator/TTree/Path            root_file_path  $ 
  $ STRING  Generator/TTree/Name                ttree_name  $
\end{minted}

\section{Algorithms for complex system studies}
\label{sec:toucans_algo}
The simulations of neutron transport can be divided into three main categories: analog, shielding and criticality studies. Analog studies aim at calculating not only average statistical properties of the nuclear system, but also look at calculating its fluctuations and correlations~\cite{Dumonteil2021}. In this case, all laws governing the transport and collision of neutrons, as well as the laws describing the nuclear fission process must be simulated analogously~\cite{Petit2017}. Variance reduction techniques that do not preserve fluctuations and correlations must therefore be inhibited and codes such as \fifrelin have to be used to reproduce the fluctuations inherent to the fission process. On the contrary, shielding studies require to use variance reduction techniques so as to provide decent scoring statistics in a context of strong attenuation. In this case, \toucans offers a plug-and-play variance reduction technique called Adaptive Multilevel Splitting (AMS), which has recently been implemented in the code. Finally, the characterization of the neutron transport in multiplicative media requires to control the neutron number at each generation, so as to prevent the extinction or the explosion of the neutron statistics. This problem is tackled with the well known power iteration algorithm (also knwon as \emph{Lanczos} algorithm), which is an iteration over generation, allowing to estimate the $k_{\text{eff}}$ of a nuclear system together with its associated eigenvector (the flux spatial distribution). Options for shielding studies and criticality studies using \toucans are described in the next subsections.

\subsection{Shielding studies and variance reduction techniques}
\label{subsec:shielding_ams}
\begin{figure}[htbp]
    \begin{center}
	\includegraphics[scale=0.4]{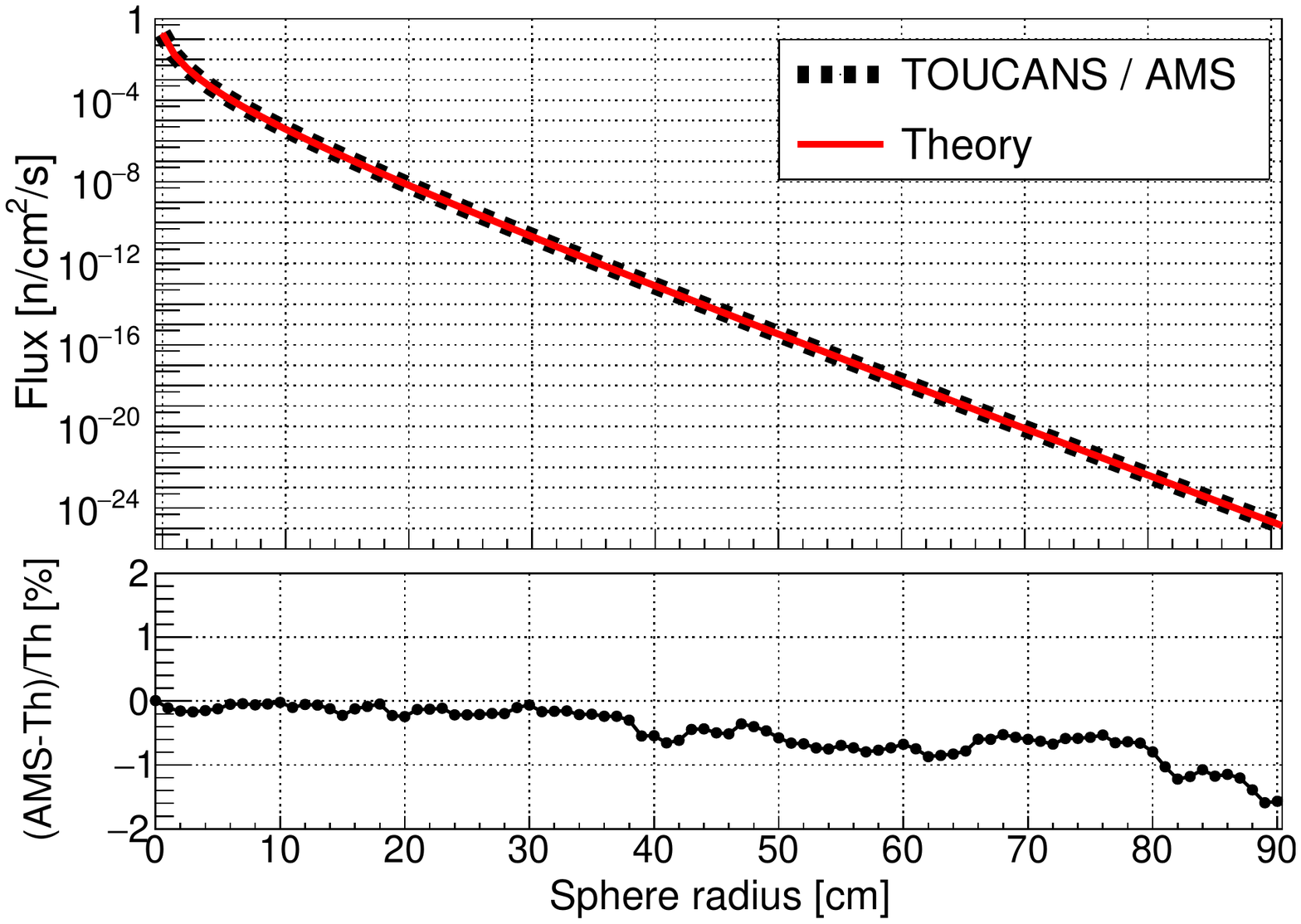} 
	\end{center}
	\caption{\label{fig:benchmark1D} Comparison between AMS in \toucans and the analytical formula for the 1D analytical benchmark.}
\end{figure}

The AMS algorithm was originally developed in the context of rare event simulation in~\cite{Cerou2007}. It has been successfully applied to neutron transport in \tripoli code by H. Louvin et al.~\cite{Louvin2017}. It consists in an iterative procedure where, at each steps, a given fraction $K/N$ of the $N$ simulated particles (with low chances to contribute to the score) is removed from the simulation and resampled at other locations (in the parameter phase space) -in such a way that no bias is introduced- by exploiting the Markovianity of the neutron transport process. To roughly estimate the chances that a particle will contribute to the score, the user provide a cost function $I$. Compared to the exponential transform approach~\cite{tripoli4} which heavily relies on the accuracy of the adjoint flux, the AMS is extremely robust and can hence be seen as a "plug-and-play" algorithm to improve the Figure Of Merit (FOM) of a simulation. The AMS algorithm has been for the first time implemented and validated in \geant through the \toucans code, which also provides a set of basic cost functions $I$ to the user (other user defined cost functions can be easily added). The main parameters as well as a simplified flowchart of the AMS algorithm are briefly summarized below: 
\begin{minted}[fontsize=\small, escapeinside=||]{C++}
///Definition
B = number of independant simulations (a.k.a. batches)
N = number of neutron histories per batch
K = number of particle removed at each iteration 
I = importance function (a.k.a. cost function)
T = splitting level chosen adaptatively during the simulation
|\textalpha| = weighting factor of the scored quantity

///Launch B independant simulations
for batch in B:

    ///Initialization
    |\textalpha| = 1

    ///Simulation
    while (N-K+1) tracks not in detector:
        1/ analog transport of the K tracks until 
        their death by capture or geometry leaking.
        2/ score assignation to the N tracks with I.
        3/ rank the N tracks as a function of their score.
        4/ define T as the Kth worst track importance.
        4/ kill/remove the K worst tracks. 
        5/ restart these K tracks in sampling one of the 
        (N-K) remaining tracks at a position just above a T.
        6/ update the weighting factor |\textalpha| at the current 
        iteration, i.e. |\textalpha| *= (1-K/N)

///Agregate the batch results, the weighted scored quantities

\end{minted}
The weight $\alpha$ is assigned to the configuration of this set of $N$ tracks, \textit{i.e.} to the scored quantity. Hence it is the same for all the particles at the final iteration of a given batch, explaining that correlations and fluctuations are preserved. The weight $\alpha$ is initially set to $1$ and multiplied by $(1-\frac{K}{N})$ at each iteration.
\noindent
The implementation of the AMS algorithm was verified with different analytical benchmarks. Here we report the result of one of these benchmarks, corresponding to the propagation of mono-directional neutrons in the one energy group approximation (the neutron energy does not change) and in nested spheres with radius ranging from 1 cm to 90 cm by 1 cm step. The neutron flux is tallied in each sub-shell. The analytical formula to compute the neutron flux in each sub-shell is given by~\cite{Placzek1953}:

\begin{equation}\label{eq:benhcmark1D}
    \phi (x) = \frac{3}{4\pi (x_{2}^{3} - x_{1}^{3})(p-1)}(e^{(p-1)\Sigma_{t}x_{2}} - e^{(p-1)\Sigma_{t}x_{1}})
\end{equation}
\noindent
where $\Sigma_{t}$ is the total macroscopic cross section, $p$=$\sigma_{el}$/$\sigma_{t}$ where  $\sigma_{el}$, $\sigma_{c}$ and $\sigma_{t}$=$\sigma_{el}$+$\sigma_{c}$ are respectively the microscopic elastic, capture and total cross-sections. In the simulation the cross-sections are set to $\Sigma_{t}$=1 cm$^{-1}$, $\sigma_{el}$=5 barns and $\sigma_{c}$=5 barns. In \geant this could has been allowed by the creation of a virtual nucleus having user defined properties, \textit{i.e.} with user defined cross-sections. 
\newline
The AMS parameters are set to $N$=10000, $K$=100, $B$=1000. The cost function is simply defined by the distance traveled by the neutron (the further, the better). The comparison between the \geant /\toucans simulation and the theory is presented in Figure~\ref{fig:benchmark1D}, exhibiting a very good agreement with differences less than 2 \% for a neutron attenuation of 10$^{+25}$. This result was obtained in only 1 hour of a mono-processor simulation, and could by no mean be calculated at all without variance reduction techniques. A paper dedicated to the AMS implementation in \geant and its extension to multi-particle/multi-detectors scorers should be released soon, along with its commit in \geant to make it available to the \geant community~\cite{Thulliez2023-2024}.

The AMS can be launched within \toucans with the following commands:
\begin{minted}[fontsize=\small]{C++}
$ BOOL        AMS/IsON                  1              $
    [0=disable (default) or 1=enable]
$ STRING      AMS/Scoring/TFile         my_output_file $
    [name of the file in whih the results are saved]
$ INTLIST     AMS/PDGCode               my_pdg_list    $
    [PDGcode of particle to follow]
$ STRING      AMS/Target/Option         my_target_opt  $
    [SINGLE_TARGET, MULTIPLE_TARGET]
$ STRINGLIST  AMS/Target/Type           my_target_type $
    [VOLUME, NEUTRON_GENERATION, TIME]
$ STRINGLIST  AMS/Target/Name           my_volume      $  
    [name of the target volume]
$ INTLIST     AMS/Target/PDGCode        my_pdg_list    $
    [PDGCode of particle to score]
$ DOUBLELIST  AMS/Target/Point          x  y  z        $ 
    [target position to push the particles to]
$ STRING      AMS/Importance/Function   my_fct         $
    [DISTANCE_POINT, DISTANCE_PLAN, NEUTRON_GENERATION, 
    TIME, IMPORTANCE_MAP, etc]
$ INTLIST     AMS/Importance/PDGCode    my_pdg         $
$ INT         AMS/Particle/Number       my_follow_nb   $
    [followed particle number/initial track number]
$ INT         AMS/ParticleToKill/Number my_kill_nb     $
    [particle to kill at each algorithm step: K]
\end{minted}

\subsection{Criticality studies - power iteration algorithm}
\label{subsec:toucans_algo_criticality}
To study critical systems, a basic power iteration algorithm has been implemented within \geant, allowing to simulate the neutron transport through fissile media (\textit{i.e.} neutron amplifier for CANS facilities) or a toy model nuclear reactor. This algorithm can be used through the following \toucans command:

\begin{minted}[fontsize=\small]{C++}
   $ BOOL       Criticality/IsON            1 $
   [0=disable (default) or 1=enable]
\end{minted}
\noindent
Basically the power iteration algorithm uses at each generation the combing algorithm~\cite{Booth1996ACarlo} to control the neutron population, \textit{i.e.} to track the same number of neutrons at each generation. Basically it chooses the neutrons that will be removed (mainly for system with $k_{\text{eff}}$>1) or duplicated (mainly for system with $k_{\text{eff}}$<1) so as to track them in the next generation. 
It is important to note that this algorithm has never been verified or validated against any international criticality benchmarks~\cite{icsbep}. Nonetheless it should constitute an interesting toy model for educational purposes and for basic criticality-safety studies such as neutron amplifiers for CANS.
The convergence of the simulation can be checked with a $k_{\text{eff}}$ or entropy criteria~\cite{Lux1991,Nowak2016}.
\newline \indent
Recently it has been shown that population control techniques could be seen as variance reduction techniques with either generational or time detectors~\cite{Frohlicher2023}. For instance, studying the properties of a sub-critical system consists in forcing the neutrons to survive and pushing them through time or generations. This is precisely the action of the AMS algorithm, which can therefore also be used to study the generational evolution of a neutron population in fissile media instead of the power iteration method. To calculate the $k_{\text{eff}}$ of a system using the AMS within \toucans, it is sufficient to use the command below. More details and examples can be found in ref~\cite{Frohlicher2023}.
\begin{minted}[fontsize=\small]{C++}
$ STRINGLIST  AMS/Target/Type   NEUTRON_GENERATION or TIME  $
\end{minted}

\section{Coupling with a multi-objective optimization algorithm, efficient design of CANS}
\label{sec:toucans_coupling}
\begin{figure}[htbp]
    \begin{center}
	\includegraphics[scale=0.35]{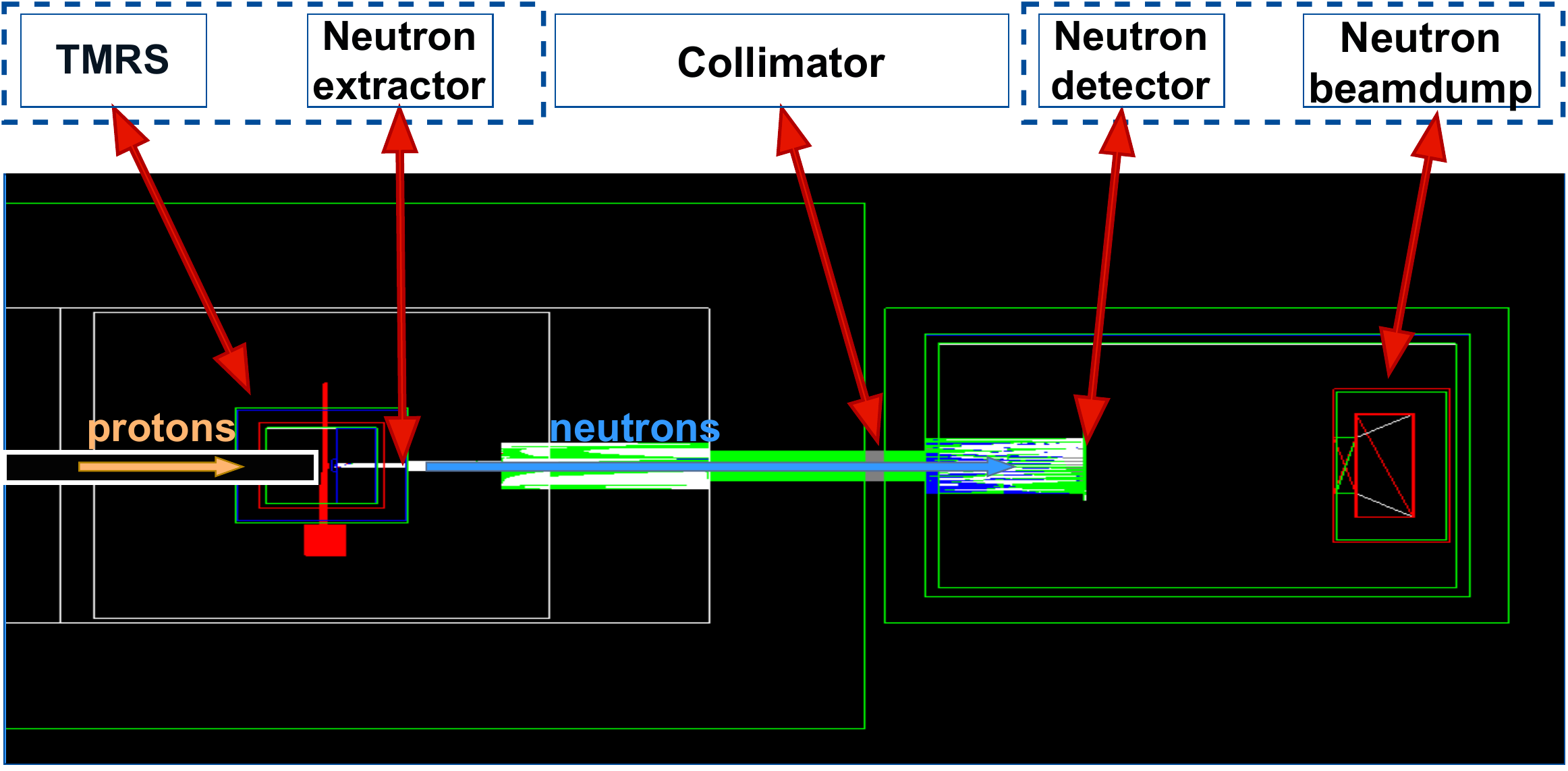}
	\end{center}
	\caption{\label{fig:cans_synoptic} Synoptic scheme of a CANS facility.}
\end{figure}

A compact accelerator based neutron source can be divided in four parts: the primary neutron source definition induced for example by (p,n) reactions, the design of the target-moderator-reflector-shielding assembly (TMRS) to slow-down the neutrons to the desired energy, the extractor/collimator to extract and transport the neutrons to the experimental point with user-defined characteristics (neutron flux, beam purity, beam divergence and spatial distribution at the detector position) and the neutron beamdump to reduce radiobiological issues along with neutron induced background on the detector. A synoptic scheme of this kind of facility is visible in Figure~\ref{fig:cans_synoptic}.

\begin{figure}[htbp]
    \begin{center}
	\includegraphics[width=.48\textwidth]{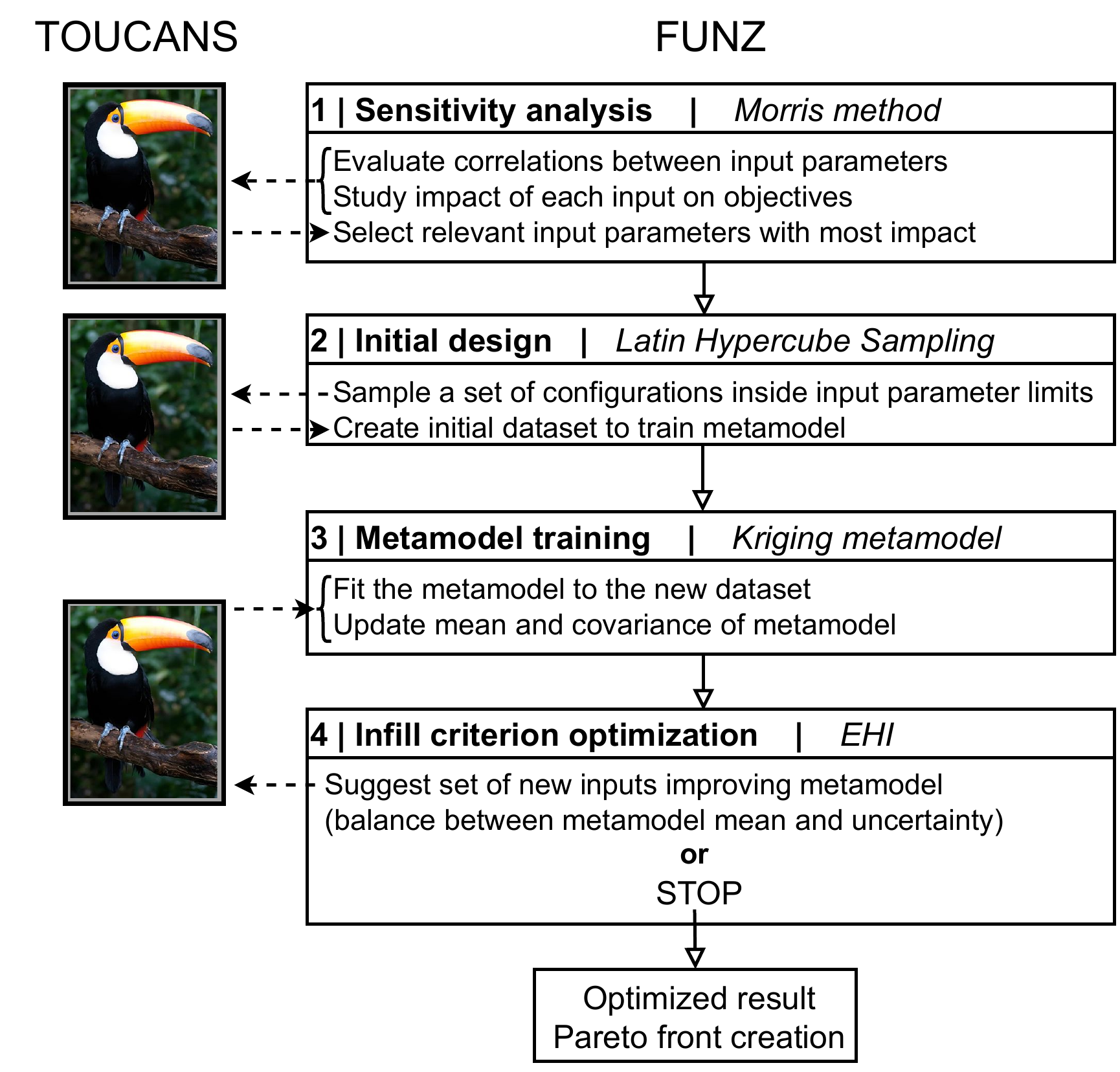}
	\end{center}
	\caption{\label{fig:toucan_promethee} Coupling scheme between \toucans which is the black-box function to discover and \funz multi-objective optimization code.}
\end{figure}

The optimization of these parts can be time consuming if traditional methods such as a search-grid algorithm are used. This limitation, often named "curse of dimensionality", appears when too many parameters effects are investigated simultaneously, which hence require evaluation strategies.
The fact that \toucans can be easily coupled to external codes thanks to its input file structure, has allowed to couple it to a multi-objective algorithm, from the GPARETO package~\cite{GPareto-CRAN,Binois2019}, using the \funz environment~\cite{Richet_Funz2021}. 
The approach consists in studying the variation of the outputs of interest, called objectives, as a function of the user-defined inputs having constraints. This allows to build a metamodel (also called surrogate model) which is a mathematical approximation of the input/output relationship of the simulation code (defined as a black-box) from which the global minimum has to be found. In this work, the black box is the Monte Carlo code \toucans which is inherently non-deterministic. The general principles of a metamodel-based optimization are described below using the Kriging metamodel~\cite{Optim_Geostatistic-10.2113/gsecongeo.58.8.1246} (also known as Gaussian process regression~\cite{ML_Rasmussen2006}). In this work the Efficient Global Optimization algorithm (EGO) (more details are given  in~\cite{Optim_EGO-Jones1998}) is used \textit{via} \funz. A flowchart of the optimization procedure is presented in Figure~\ref{fig:toucan_promethee} along with a detailed description of the steps and algorithms used to perform them. The minimization of all the objectives at once is often not possible because of conflicting interests. It is then necessary to determine the set of optimal solutions over the input parameter ranges, called a \emph{Pareto set}. The image of the \emph{Pareto set} in the objective space is called the \emph{Pareto front}, which is used to select the best solution. A detailed description of this optimization procedure is given in~\cite{Mom2022} where it has been applied to the design of a neutron beamdump to find its optimal dimensions in being in a restricted space.
	
\begin{figure}[htbp]
    \begin{center}
	\includegraphics[scale=0.25]{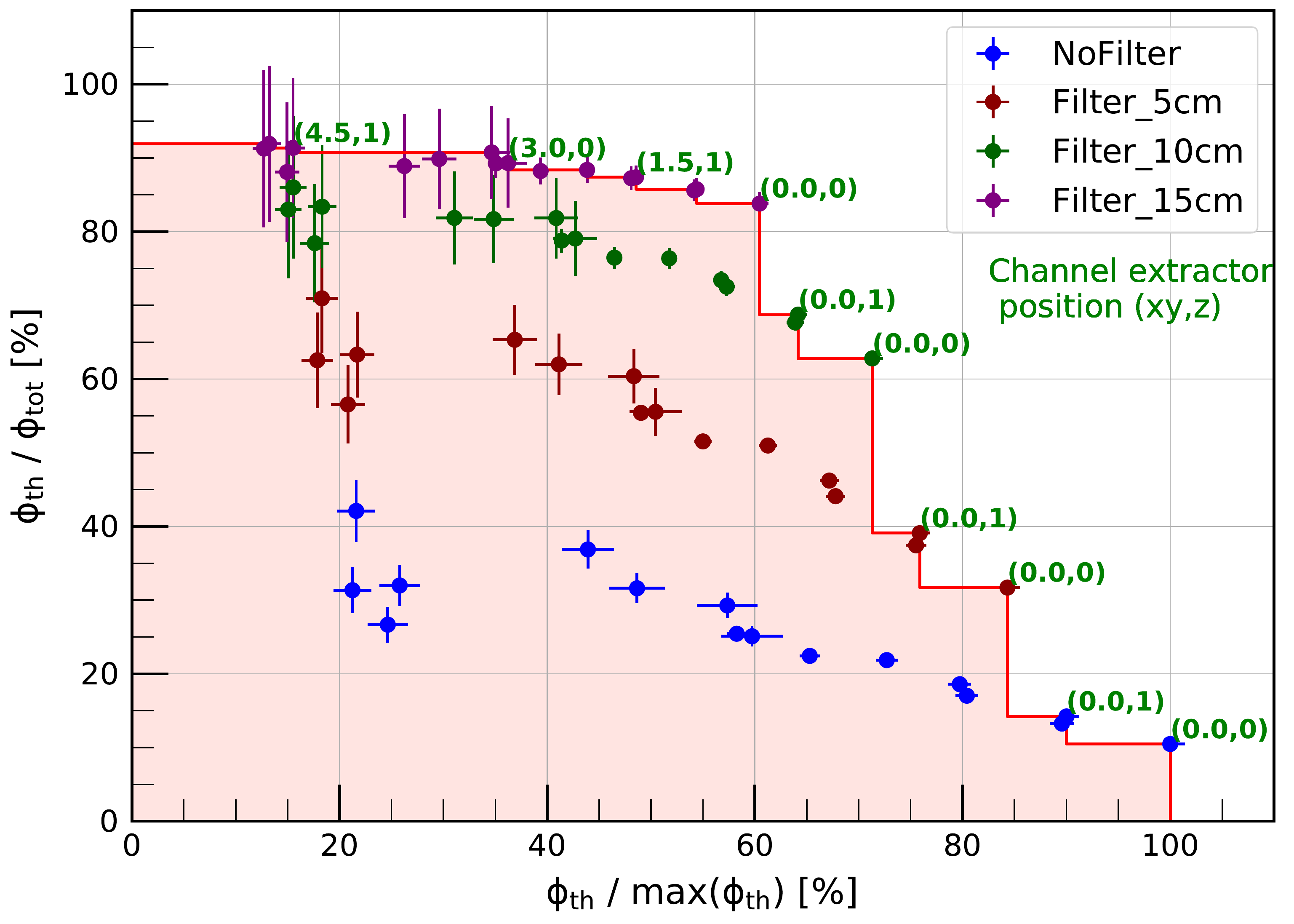}
	\end{center}
	\caption{\label{fig:pareto_extractorTMRS} Multi-objective optimization results for the extractor channel inside the TMRS assembly and single-crystal sapphire filter.}
\end{figure}

In this work, the EGO algorithm has been applied to the neutron extractor channel configuration in the TMRS assembly associated with a single-crystal sapphire filter which is used as a thermal neutron pass filter. The objective is to simultaneously maximize the thermal neutron flux ($\phi_{th}$) as well as the ratio of the thermal neutron flux over the total flux ($\frac{\phi_{th}}{\phi_{tot}}$). The results of this optimization is given in Figure~\ref{fig:pareto_extractorTMRS}, where each color is associated with a filter length and for each of them, different positions of the extractor channel in the TMRS are evaluated. The thermal neutron flux is maximum when the extractor channel is close to the neutron production location and decreases when it is moved away. In Figure~\ref{fig:pareto_extractorTMRS} the bold red line represents the \emph{Pareto front} which is the set of optimal configuration points from which the decision maker can choose to achieve the best trade-off.

\section{Conclusion}	
Initially designed to support CANS design activities at CEA-Saclay and fundamental physics experiments requiring a good accuracy concerning the transport of neutrons (\textit{e.g.} the CRAB experiment~\cite{Thulliez2021,crab2022}), the \toucans code leverages on recent improvements of the \geant Neutron-HP package. Through its numerous implicit and external coupling to other codes such as \fifrelin, \ncrystal, \cadmesh or \funz, \toucans allows to simulate and optimize in a single framework all the aspects of an experimental setup, from the neutron production to the detector response, and can therefore be used for shielding or criticality-safety studies, neutron noise estimation or nuclear instrumentation just to name a few. In a close future, the code should be released as one of \geant advanced examples.

\section*{Acknowledgement}

The authors would like to thank Antoine Drouart (CEA-Saclay, France) for fruitful discussions about CANS, Olivier Litaize (CEA-Cadarache, France) and Yann Richet (IRSN, France) for their support on all aspects, respectively, related to \fifrelin and \funz.

\bibliographystyle{unsrtnat}
\bibliography{./references}.

\appendix
\section{Appendix}	
\label{annexe:inputCard}

\begin{figure}[htbp]
    \begin{center}
	\includegraphics[scale=0.4]{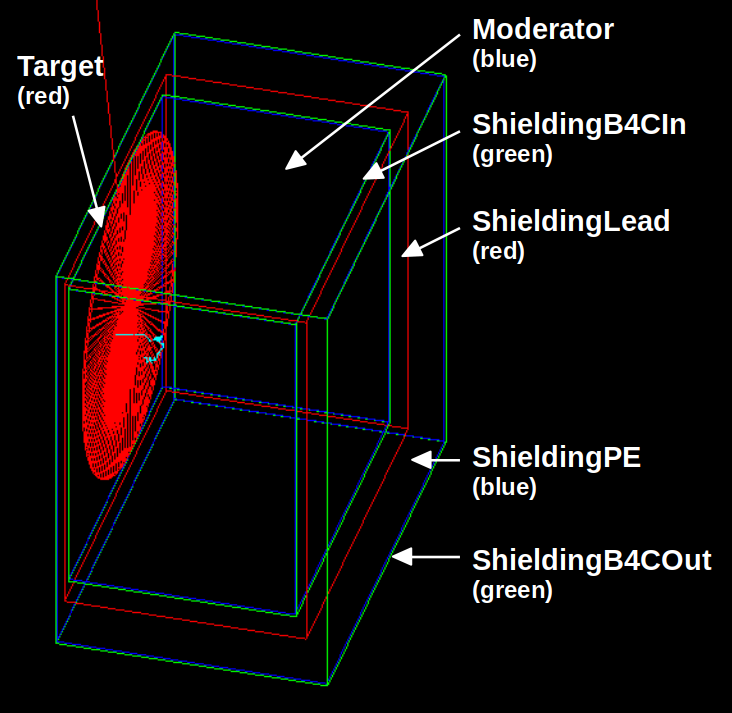}
	\end{center}
	\caption{\label{fig:annexe_toucans_visu} \geant visualization of the \toucans input file presented in appendix.}
\end{figure}

Example of very simple input file with its main sections (geometry, source, tallies) and commentaries. The corresponding visualization is presented in Figure~\ref{fig:annexe_toucans_visu}. This input file allows to simulate a simple moderator placed after a proton-to-neutron converter wrapped around in a radio-biological shielding.

\begin{minted}[fontsize=\small]{C++}
//***************************************** 
// TOUCANS input file
//***************************************** 

$ BOOL      Geant4/Visualization                1                $ 
$ STRING    Geant4/Macro                        USER_INPUT_FILE  $ 
$ STRING    Geant4/OutputFileName               USER_OUTPUT_FILE $ 
$ INT       Geant4/ParticleToShoot              50000            $
$ STRING    Geant4/PhysicsList                  QGSP_BERT_EMZ    $ 
$ INT       Geant4/Seed/Index                   1156             $ 
$ BOOL      Geant4/StepLimiter/Instantiation    0                $ 
$ DOUBLE    Geant4/Cut/Energy                   0                $  
$ DOUBLE    Geant4/Cut/Range                    0.               $  
$ BOOL      Geant4/CheckOverlaps                1                $ 
$ BOOL      Geant4/SteppingAction/Print         0                $ 
$ BOOL      Geant4/InteractiveSession           0                $ 
$ BOOL      Geant4/ExportGeometryGDML           0                $

//***************************************** 
// Path to data needed by the simulation
//***************************************** 
$ STRING        Data/Directory          USER_DATA_DIRECTORY_PATH $

//***************************************** 
// Mapping
//***************************************** 
$ STRINGLIST    Geant4/VirtualMapping/List  MapNeutron  MapGamma $
 
$ DOUBLELIST    ScoringMesh0/Dimensions      1000 1000  1500     $
$ INTLIST       ScoringMesh0/Binning         100  100   150      $
$ DOUBLELIST    ScoringMesh0/Position        0.  0. 0.           $

$ STRING        MapNeutron/Mesh                  ScoringMesh0    $
$ STRING        MapNeutron/Name                  mapNeutron      $
$ STRING        MapNeutron/Scorer                cellFlux        $
$ STRINGLIST    MapNeutron/Filter/Particle       neutron         $
$ STRINGLIST    MapNeutron/Filter/EnergyDomain   1E-11  100E-6   $

$ STRING        MapGamma/Mesh                  ScoringMesh0      $
$ STRING        MapGamma/Name                  mapGamma          $
$ STRING        MapGamma/Scorer                cellFlux          $
$ STRINGLIST    MapGamma/Filter/Particle       gamma             $
$ STRINGLIST    MapGamma/Filter/EnergyDomain   1E-6   20         $

//***************************************** 
// Primary generator
//***************************************** 
$ INTLIST     SteppingAction/ParticleToFollow/PDGCode    2112 22 $ 
    [only neutrons and gammas are transported in the simulation]
    
$ INT         PrimaryGenerator/PDGCode                      2112 $ 
    [emitted particle type]

$ STRING      PrimaryGenerator/Type     ENERGY_ANGLE_CORRELATION $ 
///trunk=PrimaryGenerator  
$ STRING      trunk/EnergyAngleSampler            TH2F           $ 
$ STRING      trunk/EnergyAngleSampler/TH2F/Path  USER_ROOT_FILE $ 
$ STRING      trunk/EnergyAngleSampler/TH2F/Name  USER_HIST_NAME $ 

$ STRING      PrimaryGenerator/PositionSampler    GAUSSIAN       $ 
///trunk=PrimaryGenerator/PositionSampler 
$ DOUBLE      trunk/Gaussian/SigmaX               1.2            $ 
$ DOUBLE      trunk/Gaussian/SigmaY               0.8            $ 
$ DOUBLE      trunk/ZDepth                        0.02           $ 
$ DOUBLELIST  trunk/Global/Position               0. 0. 0.       $ 

$ STRING      PrimaryGenerator/MomentumSampler  RESTRICTED_THETA $ 
///trunk=PrimaryGenerator/MomentumSampler
$ DOUBLE      trunk/RestrictedTheta/Theta         0             $ 
$ STRINGLIST  trunk/Global/Rotation/Axis          X             $ 
$ DOUBLELIST  trunk/Global/Rotation/Angle         0             $ 

//***************************************** 
// Geomtry - World definition
//***************************************** 
$ STRING        World/Type                      BOX             $ 
$ STRING        World/Material                  G4_Galactic     $
$ DOUBLELIST    World/Position                  0. 0. 0.        $
$ DOUBLELIST    World/Dimensions                3000 3000 3000  $
$ STRING        World/Color                     black           $
  
 
//***************************************** 
// Geometry - User own geometry 
//***************************************** 
$ STRING        Geometry/Instantiation/Type         FULL_USER   $ 
$ STRINGLIST    Object/List	Target  ShieldingB4COuter 
                                    ShieldingPE 
                                    ShieldingLead 
                                    ShieldingB4CInner 
                                    Moderator                   $ 

$ STRING        Target/Type                     CYLINDER        $ 
$ STRING        Target/Material                 GRAPHITE        $ 
$ STRING        Target/MotherVolume             World           $
$ STRING        Target/Color                    red             $
$ STRINGLIST    Target/Rotation/Axis            X               $
$ DOUBLELIST    Target/Rotation/Angle           0.              $
$ DOUBLELIST    Target/Dimensions               0 60 0.7        $  
    [Rmin, Rmax, Z]
$ DOUBLELIST    Target/Position                 20. 0. -0.35    $ 

$ STRING        ShieldingB4COut/Type          BOX               $
$ STRING        ShieldingB4COut/Material      BOREFLEX          $ 
$ STRING        ShieldingB4COut/MotherVolume  World             $
$ STRING        ShieldingB4COut/Color         green             $
$ DOUBLELIST    ShieldingB4COut/Dimensions    152 152 96        $  
    [X, Y, Z]
$ DOUBLELIST    ShieldingB4COut/Position      0. 0. 48          $ 

$ STRING        ShieldingPE/Type                BOX             $ 
$ STRING        ShieldingPE/Material            POLYETHYLENE    $ 
$ STRING        ShieldingPE/MotherVolume        ShieldingB4COut $
$ STRING        ShieldingPE/Color               blue            $
$ DOUBLELIST    ShieldingPE/Dimensions          151 151 95.5    $
$ DOUBLELIST    ShieldingPE/Position            0. 0. -0.25     $

$ STRING        ShieldingLead/Type              BOX             $ 
$ STRING        ShieldingLead/Material          G4_Pb           $ 
$ STRING        ShieldingLead/MotherVolume      ShieldingPE     $
$ STRING        ShieldingLead/Color             red             $
$ DOUBLELIST    ShieldingLead/Dimensions        131 131 85.5    $
$ DOUBLELIST    ShieldingLead/Position          0. 0. -5        $

$ STRING        ShieldingB4CIn/Type             BOX             $ 
$ STRING        ShieldingB4CIn/Material         BOREFLEX        $ 
$ STRING        ShieldingB4CIn/MotherVolume     ShieldingLead   $
$ STRING        ShieldingB4CIn/Color            green           $
$ DOUBLELIST    ShieldingB4CIn/Dimensions       121 121 80.5    $
$ DOUBLELIST    ShieldingB4CIn/Position         0. 0. -2.5      $    

$ STRING        Moderator/Type                  BOX             $
$ STRING        Moderator/Material              LIGHT_WATER     $ 
$ STRING        Moderator/MotherVolume          ShieldingB4CIn  $
$ STRING        Moderator/Color                 blue            $
$ STRINGLIST    Moderator/Rotation/Axis         X               $
$ DOUBLELIST    Moderator/Rotation/Angle        0.              $
$ DOUBLELIST    Moderator/Dimensions            120 120 80      $
$ DOUBLELIST    Moderator/Position              0. 0. -0.25     $ 
$ STRINGLIST    Moderator/Scorers               CELL_FLUX       $ 
\end{minted}

\end{document}